\documentstyle[12pt,epsf,psfig]{article}

\newcommand{\lsim}{\mathrel{\raisebox{-.6ex}{$\stackrel{\textstyle<}{\sim}$}}}

\def\beq{\begin{equation}}
\def\eeq{\end{equation}}
\def\bea{\begin{eqnarray}}
\def\eea{\end{eqnarray}}

\begin{document}

\thispagestyle{empty}

\font\fortssbx=cmssbx10 scaled \magstep2
\hbox to \hsize{
%\special{psfile=/NextLibrary/TeX/tex/inputs/uwlogo.ps
%                         hscale=8000 vscale=8000
%                         hoffset=-12 voffset=-2}
%\hskip.5in\raise.1in
%\hbox{\fortssbx University of Wisconsin - Madison}
      \hfill$\vtop{
\hbox{MADPH-02-1254}
\hbox{AMES-HET-02-01}}
$}

\vspace{.5in}

\begin{center}
{\large\bf Neutrinoless double beta decay can constrain \\ neutrino dark matter}\\
\vskip 0.4cm
{V. Barger$^1$, S.L.~Glashow$^2$, D. Marfatia$^2$, and K. Whisnant$^3$}
\\[.1cm]
$^1${\it Department of Physics, University of Wisconsin, Madison, WI
53706, USA}\\
$^2${\it Department of Physics, Boston University, Boston, MA 02215, USA}\\
$^4${\it Department of Physics and Astronomy, Iowa State University,
Ames, IA 50011, USA}\\
\end{center}

\vspace{.5in}

\begin{abstract}                                    

We examine how constraints can be placed on the neutrino component of
dark matter by an accurate measurement of neutrinoless double beta 
($0\nu\beta\beta$) decay and the solar oscillation amplitude.
We comment on the alleged evidence for $0\nu\beta\beta$ decay.

\end{abstract}

\thispagestyle{empty}
\newpage

The detection of neutrinoless double beta decay would imply the violation of
lepton number conservation. The process could be induced by Majorana neutrino
mass terms, or by less trivial modifications of the standard model.
Here we consider the former possibility wherein there are exactly three
left-handed neutrino states with Majorana masses~\cite{work}.
The  measurement of $0\nu\beta\beta$ decay, together with what has been
learned from studies of solar and atmospheric neutrinos, 
has direct consequences on the spectrum of neutrino masses and therefore on
the effects of neutrinos on structure formation. We define what will
be necessary to determine the neutrino component of dark matter 
from terrestrial
experiments. 

The charged-current eigenstates are related to the
mass eigenstates by a unitary transformation
\beq
\left( \begin{array}{c} \nu_e \\ \nu_\mu \\ \nu_\tau \end{array} \right)
= U V \left( \begin{array}{c} \nu_1 \\ \nu_2 \\ \nu_3 \end{array} \right)
= \left( \begin{array}{ccc}
  c_{2} c_{3}                           & c_{2} s_{3}
& s_{2} e^{-i\delta} \\
- c_{1} s_{3} - s_{2} s_{1} c_{3} e^{i\delta} &   c_{1} c_{3} - s_{2} s_{1} s_{3} e^{i\delta}
& c_{2} s_{1} \\
  s_{1} s_{3} - s_{2} c_{1} c_{3} e^{i\delta} & - s_{1} c_{3} - s_{2} c_{1} s_{3} e^{i\delta}
& c_{2} c_{1} \\
\end{array} \right) V
\left( \begin{array}{c}
\nu_1 \\ \nu_2 \\ \nu_3
\end{array} \right) \,,
\label{eq:U}
\eeq
where $s_i$ and $c_i$ are the sines and cosines of $\theta_i$,
and $V$ is the diagonal matrix 
$(1,e^{i{\phi_2 \over 2}},e^{i({\phi_3\over 2}+\delta)}$).
In Eq.~(\ref{eq:U}), $\phi_2$ and $\phi_3$ are additional phases for Majorana
neutrinos that are not measurable in neutrino oscillations;
if $CP$ is conserved, the phases in $UV$ are either $0$ or $\pi$.

The solar neutrino data favor the Large Mixing Angle solution with
$0.6 \le \sin^22\theta_{3} \le 0.98$, and $2 \times
10^{-5}$~eV$^2 \le \Delta_s \le 4 \times 10^{-4}$~eV$^2$ at the 
3$\sigma$ C.L.~\cite{Bahcall:2001cb}.
Atmospheric neutrino data imply $\sin^22\theta_{1} \ge 0.85$ and
$ 1.1 \times 10^{-3}$~eV$^2 \le \Delta_a \le 5 \times
10^{-3}$~eV$^2$ at the 99\% C.L.~\cite{Toshito:2001dk}. 
The CHOOZ reactor experiment imposes
the constraint $\sin^22\theta_{2} \le 0.1$ at the 95\% C.L.~\cite{CHOOZ}.
$\Delta_s$ and $\Delta_a$ are the 
mass-squared differences relevant to
solar and atmospheric neutrino oscillations, respectively.

We choose the mass ordering $m_1 < m_2 < m_3$ with $m_i$ non-negative. 
There are two possible
neutrino mass spectra:
\begin{eqnarray}
\Delta_s=m_2^2-m_1^2\,,&& \Delta_a=m_3^2-m_2^2\,,\ \ \ \ \ \ {\rm (normal\
  hierarchy),}\\
\Delta_s=m_3^2-m_2^2\,,&& \Delta_a=m_2^2-m_1^2\,,\ \ \ \ \ \ {\rm (inverted\
  hierarchy),}
\end{eqnarray}
where in either case $\Delta_a \gg \Delta_s$ in accord with the previously
described experimental data.
 For the {\it normal hierarchy} (Case~I), mixing 
is given by
Eq.~(\ref{eq:U}).  The limit on $\theta_{2}$  implies
that there is very little mixing of $\nu_e$ with the heaviest state.
In Case~I solar neutrinos oscillate  primarily between the
two lighter mass eigenstates. 
  For the  {\it inverted hierarchy} (Case~II),  solar neutrinos
oscillate primarily between the two nearly degenerate 
heavier states. In this case
the mixing is described  by interchanging the roles of
$m_1$ and $m_3$. With a  mixing matrix
 obtained from Eq.~(\ref{eq:U}) by
interchange of the first and third columns of $UV$, the parameters
governing neutrino oscillations ($\theta_i$ and $\delta$) retain
 the same import 
as those in Case~I. The limit on $\theta_{2}$ again
implies that for  Case~II 
there is very little mixing of $\nu_e$ with the lightest
state.

The rate of $0\nu\beta\beta$ decay depends on the magnitude of
the $\nu_e$--$\nu_e$ element of the neutrino mass matrix~\cite{0nubb},
which is
\bea
M_{ee} &=& |c_{2}^2 c_{3}^2 m_1 + c_{2}^2 s_{3}^2 m_2 e^{i\phi_2}
+  s_{2}^2 m_3 e^{i\phi_3}| \,, \ \ \ {\rm~(Case~I)} \,,
\label{Mee1}\\
&=& |c_{2}^2 c_{3}^2 m_3 + c_{2}^2 s_{3}^2 m_2 e^{i\phi_2}
+  s_{2}^2 m_1 e^{i\phi_3}| \,, \ \ \ {\rm~(Case~II)} \,.
\label{Mee2}
\eea
The masses $m_i$ may be determined from the lightest mass $m_1$ and the
mass--squared differences. Since the solar mass--squared difference is
very small it can be ignored; then setting $m_1=m$ and $\Delta_a=\Delta$,
\bea
&& m_2 = m \,, \qquad m_3 = \sqrt{m^2 + \Delta} \,,
\ \ \  {\rm~(Case~I)} \,,
\label{eq:mass1}\\
&& m_2 = m_3  =  \sqrt{m^2 + \Delta} \,,
\ \ \ \ \ \ \ \ \ \ \ \ \ \,{\rm~(Case~II)} \,.
\label{eq:mass2}
\eea
The lightest mass is related to the sum of neutrino masses ($\Sigma = \Sigma m_i$) 
via
\bea
\Sigma &=& 2 m + \sqrt{m^2 + \Delta} \,,
\ \ \ {\rm~(Case~I)} \,,
\label{eq:sum1}\\
\Sigma &=& m + 2\sqrt{m^2 + \Delta} \,,
\ \ \ {\rm~(Case~II)} \,.
\label{eq:sum2}
\eea

For a given value of  $M_{ee}$, the minimum possible value of $m$ 
is obtained if the three contributions to
$M_{ee}$ are  in phase, {\it i.e.,}  $\phi_2 = \phi_3 = 0$. Thus
\bea
m_{\rm min} &=&
{M_{ee} c_{2}^2 - s_{2}^2 \sqrt{M_{ee}^2 +\Delta 
\cos2\theta_{2}} \over \cos2\theta_{2}} \,,
\ \ \ {\rm~(Case~I)} \,,
\label{eq:min1}\\
&=& {c_{2}^2 \sqrt{M_{ee}^2 - \Delta\cos2\theta_{2}} -
s_{2}^2 M_{ee} \over \cos2\theta_{2}} \,,
\ \ \ {\rm~(Case~II)} \,.
\label{eq:min2}
\eea
The maximum possible value of $m$
is obtained if the  the two smaller
contributions to $M_{ee}$ are out of phase with the largest
contribution ({\it i.e., } $\phi_2 = \phi_3 = \pi$ when $c_{3} > s_{3}$).
Then
\bea
m_{\rm max} &=&
{M_{ee}c_{2}^2|\cos2\theta_{3}| + s_{2}^2 \sqrt{M_{ee}^2 +
\Delta (c_{2}^4\cos^22\theta_{3}-s_{2}^4)} \over
c_{2}^4\cos^22\theta_{3}-s_{2}^4} \,, \ \ \ {\rm~(Case~I)} \,,
\label{eq:max1}\\
&=& {M_{ee} s_{2}^2 + c_{2}^2|\cos2\theta_{3}| \sqrt{M_{ee}^2 -
\Delta (c_{2}^4\cos^22\theta_{3}-s_{2}^4)} \over
c_{2}^4\cos^22\theta_{3}-s_{2}^4} \,, \ \ \ {\rm~(Case~II)} \,.
\label{eq:max2}
\eea
The allowed ranges for $\Sigma$ are  determined from
Eqs.~(\ref{eq:sum1}) and (\ref{eq:sum2}).
Because  $\theta_{2}$ is small ($\sin^22\theta_{2} \le 0.1$ or $s_{2}^2
\le 0.026$), its value does not significantly  affect the
result. (We have confirmed this result numerically).
The limits on $\Sigma$ (for $\theta_{2} = 0$) are
\bea
2 M_{ee} + \sqrt{M_{ee}^2 \pm \Delta} \le &\Sigma& \le
{2 M_{ee} + \sqrt{M_{ee}^2 \pm \Delta\cos^22\theta_{3}}
\over |\cos2\theta_{3}|} \,, 
\label{eq:sumlim1}
\eea
where the plus sign applies to the normal hierarchy and the minus sign
to the inverted hierarchy.
The bounds depend on only two oscillation parameters: the scale of
atmospheric neutrino oscillations
($\Delta$) and the amplitude of solar neutrino oscillations 
($\sin^2{2\theta_3}$).

\begin{figure}[ht]
\centering\leavevmode
\mbox{\psfig{file=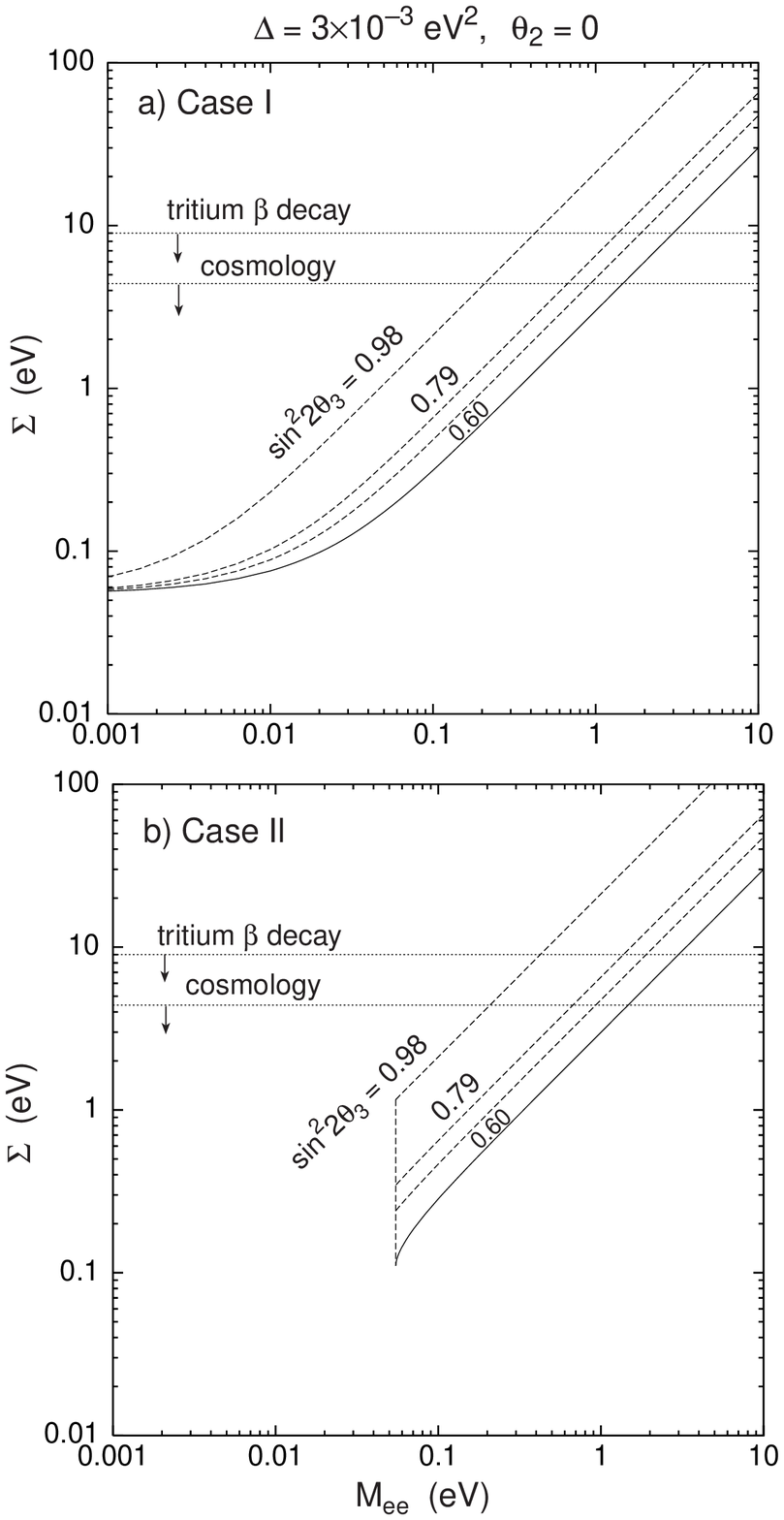,width=9cm,height=15cm}}
\medskip
\caption{$\Sigma$ versus $M_{ee}$ for the a) normal hierarchy and b) 
inverted hierarchy. The solid line is the $\theta_3$-independent lower bound on
$\Sigma$ and the dashed lines are the upper bounds for different values of 
$\theta_3$. For the inverted hierarchy, $M_{ee}\geq \sqrt{\Delta}$. The 95\% C.L.
bounds from tritium beta decay~\cite{Groom:in} and cosmology~\cite{wang} are shown.}
\label{fig}
\end{figure}

Figure~\ref{fig} shows the allowed bands for $\Sigma$ and $M_{ee}$ with several
possible  values
of $\theta_3$ within its 3$\sigma$ allowed range. 
We have fixed $\Delta=3\times 10^{-3}$ eV$^2$ and $\theta_2=0$.
The solid line is the $\theta_3$-independent lower bound on $\Sigma$ from 
Eq.~(\ref{eq:sumlim1}). The several dashed lines
 (labelled by $\sin^2{2\theta_3}$) are upper bounds on $\Sigma$.  
The bands between  solid and dashed lines
are  the ($\theta_3\,$-dependent) allowed domains  of $\Sigma$ and $M_{ee}$.
The bands in Fig.~\ref{fig}b terminate at $M_{ee}=0.055$ eV because 
$M_{ee}\geq \sqrt{\Delta}$ for Case~II.  It is possible that CP violation in
the neutrino sector is absent or negligible. In this case, the point defined
by $\Sigma$ and $M_{ee}$ must lie on one of the bounding lines of the allowed 
region.
%for a given value of
%$M_{ee}$, $\Sigma$ must lie on the solid line.

If the recent evidence that $0.05$ eV $\leq M_{ee} \leq 0.84$ eV 
at the 95\% C.L.~\cite{Klapdor-Kleingrothaus:2002ke} 
is borne out, this would imply that $0.1$ eV $\lsim \Sigma \lsim 20$ eV, 
which using $\Omega_\nu h^2= \Sigma/(93.8\, {\rm{eV}})$ translates to
\bea
0.001 \lsim \Omega_\nu h^2 \lsim 0.2\,,
\label{eq:bound}
\eea
where $\Omega_\nu$ is the fraction of the critical density contributed by neutrinos 
and $h$ is the dimensionless Hubble constant ($H_0=100h$ km s$^{-1}$ Mpc$^{-1}$). 
 CMB measurements and galaxy cluster surveys 
already constrain $\Sigma$ to be
smaller than 4.4 eV ($\Omega_\nu h^2 \lsim 0.05$) at the 95\% C.L.~\cite{wang}. 
Data from the MAP satellite
should  either determine $\Sigma$ or 
 tighten this constraint to about 0.5 eV in the near 
future~\cite{Eisenstein:1998ki}. A more stringent 
upper bound on $\Sigma$ from terrestrial experiments 
must await the precise determination of 
$\theta_3$
 (such as is anticipated from KamLAND~\cite{barger}), 
and a firmer measurement (or constraint) on $M_{ee}$~\cite{moreexps}. 

\vskip 0.4in
\noindent
{\it Acknowledgements.}  This work was supported in part by the NSF under
grant No.~NSF-PHY-0099529, in part by the
U.S. Department of Energy under grant
Nos.~DE-FG02-91ER40676, 
~DE-FG02-95ER40896 and ~DE-FG02-01ER41155, and in part by the
Wisconsin Alumni Research Foundation.

\clearpage

\end{document}